\documentclass[sigconf]{acmart}

\usepackage{xspace}
\usepackage{amsthm}
\usepackage{amsmath}
\usepackage{bm}
\usepackage{enumitem}
\usepackage{booktabs}
\usepackage{multirow}
\usepackage{graphicx}
\usepackage[normalem]{ulem}
\useunder{\uline}{\ul}{}
\usepackage{caption}
\usepackage{subcaption}

\newcommand{\modelname}{\textsf{ASReP}\xspace}
\newcommand{\transformer}{\textsf{Trm}\xspace}

\setlist[itemize]{leftmargin=*}
\AtBeginDocument{%
  \providecommand\BibTeX{{%
    \normalfont B\kern-0.5em{\scshape i\kern-0.25em b}\kern-0.8em\TeX}}}

\copyrightyear{2021}
\acmYear{2021}
\setcopyright{acmcopyright}\acmConference[SIGIR '21]{Proceedings of the 44th
International ACM SIGIR Conference on Research and Development in Information
Retrieval}{July 11--15, 2021}{Virtual Event, Canada}
\acmBooktitle{Proceedings of the 44th International ACM SIGIR Conference on
Research and Development in Information Retrieval (SIGIR '21), July 11--15, 2021,
Virtual Event, Canada}
\acmPrice{15.00}
\acmDOI{10.1145/3404835.3463036}
\acmISBN{978-1-4503-8037-9/21/07}

\begin{document}

\title{Augmenting Sequential Recommendation with Pseudo-Prior Items via Reversely Pre-training Transformer}

\author{Zhiwei Liu*, Ziwei Fan*, Yu Wang, Philip S. Yu}\thanks{*Both authors contribute equally.}
% \email{zliu213@uic.edu}
\affiliation{%
  \institution{Department of Computer Science, University of Illinois at Chicago}
}
\email{{zliu213,zfan20,ywang617,psyu}@uic.edu}

% \author{Zhiwei LIu*, Ziwei Fan*}
% \thanks{*contribut}
% \author{Zhiwei}\authornote{Both}
% \author{Ziwei Fan}\authornotemark[1]
% \email{zliu213@uic.edu}
% \author{Yu Wang}
% \email{ywang617@uic.edu}
% \affiliation{%
%   \institution{Department of Computer Science, University of Illinois at Chicago}
% }
% \author{Ziwei Fan}
% \author{Yu Wang}
% \author{Philip S. Yu}
% \affiliation{%
%   Department of Computer Science,\\University of Illinois at Chicago
%  }
% \email{{zliu213,zfan20,ywang617,psyu}@uic.edu}

\renewcommand{\shortauthors}{Liu, etc}

\fancyhead{}

\begin{abstract}
% Sequential Recommendation characterizes the evolving patterns of users' historical transaction records by modeling item sequences chronologically. The essential target of it is to capture the item transition correlations in sequences. The recent developments of transformer inspire the community to design effective sequence encoders, \textit{e.g.,} SASRec and BERT4Rec. However, we observe these transformer-based models suffer from the cold-start issue, \textit{i.e.,} performing poorly for short sequences. Therefore, we propose to augment short sequences while still preserving original sequential correlations. We introduce a new framework for \textbf{A}ugmenting \textbf{S}equential \textbf{Re}commendation with \textbf{P}seudo-prior items~(\modelname). We firstly pre-train a transformer with sequences in a reverse direction to predict prior items. Then, we use this transformer to generate fabricated historical items, i.e., pseudo-prior items, at the beginning of short sequences. Finally, we fine-tune the transformer using these augmented sequences from the time order to predict the next item. Experiments on two real-world datasets verify the effectiveness of \modelname. Empirical results and detailed analyses demonstrate the necessity of augmenting short sequences via \modelname. The code is available on \url{https://github.com/DyGRec/ASReP}
Sequential Recommendation characterizes the evolving patterns by modeling item sequences chronologically. The essential target of it is to capture the item transition correlations. The recent developments of transformer inspire the community to design effective sequence encoders, \textit{e.g.,} SASRec and BERT4Rec. However, we observe that these transformer-based models suffer from the cold-start issue, \textit{i.e.,} performing poorly for short sequences. Therefore, we propose to augment short sequences while still preserving original sequential correlations. We introduce a new framework for \textbf{A}ugmenting \textbf{S}equential \textbf{Re}commendation with \textbf{P}seudo-prior items~(\modelname). We firstly pre-train a transformer with sequences in a reverse direction to predict prior items. Then, we use this transformer to generate fabricated historical items at the beginning of short sequences. Finally, we fine-tune the transformer using these augmented sequences from the time order to predict the next item. Experiments on two real-world datasets verify the effectiveness of \modelname. The code is available on \url{https://github.com/DyGRec/ASReP}.
\end{abstract}







\maketitle

\section{Introduction}
Recommender systems, which predict the potential interests of users to items, are widely applied in online platforms~\cite{pinsage2018ying,liu2019jscn,liu2020basconv,liu2020basket,yang2021consisrec,zhou2020gan} nowadays. 
% Owing to the dynamic characteristics of users' behavior, it is necessary to characterize the evolving patterns of users' historical records~\cite{xiong2010temporal,kang2018self,kumar2019predicting,ye2020time}.
Because of the dynamic characteristics of users' behavior, characterizing the evolving patterns of users' historical records is necessary~\cite{xiong2010temporal,kang2018self,kumar2019predicting,ye2020time,li2020dynamic,zhou2018unlearn}. Among them, the Sequential Recommendation~(SR)~\cite{rendle2010factorizing,kang2018self,zheng2019gated,li2020time,sun2019bert4rec} is studied a lot for decades. It models the sequential patterns of users' transactions on items. Therefore, the next item can be inferred.

The essential target in SR is to capture the item transition correlations~\cite{luo2019hybrid}. The recent developments of transformer~\cite{vaswani2017attention,zhang2020graph} provide a powerful backbone to embed sequence, which can effectively model item correlations. SASRec~\cite{kang2018self} is the pioneering work adopting transformer to complete SRs. It employs a dot-product self-attention module to learn the importance of items at different positions in a sequence. BERT4Rec~\cite{sun2019bert4rec}, which is inspired by bi-direction transformer~\cite{devlin2018bert}, models the item transition correlations from both left-to-right and right-to-left directions. Other recent works~\cite{ssept20wu,li2020time,ren2020sequential} also justify the efficacy of transformer in revealing item correlations in sequences.

\begin{figure}
    \centering
    \includegraphics[width=0.45\textwidth]{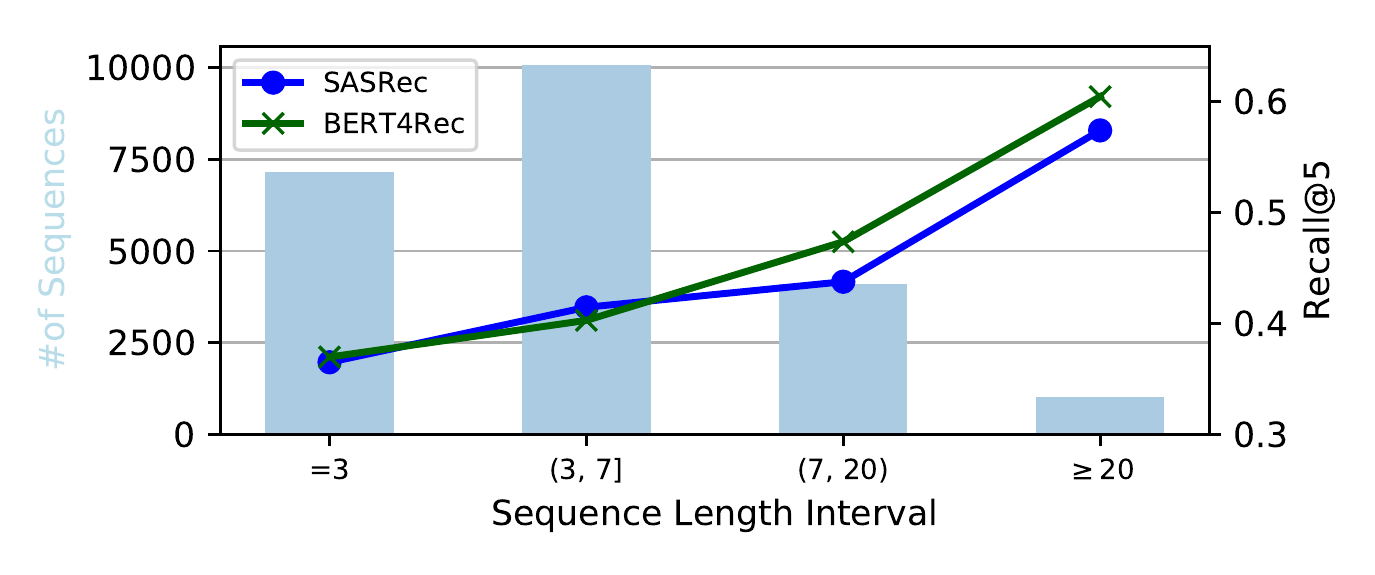}
    \vspace{-5mm}
    \caption{The sequence length distribution (bar) and the corresponding Recall@5 (line) on Amazon Beauty dataset.}
    \label{fig:intro_sasrec_perf_seqlen_dist}
\end{figure}

However, transformer-based sequence encoders fail to achieve satisfying performance when sequences are very short, \textit{i.e.,} the \textit{cold-start} issue~\cite{li2019zero}.  We illustrate the sequence length distribution and the corresponding  Recall@$5$ for next item prediction on Amazon Beauty dataset~\cite{mcauley2015image} in Figure~\ref{fig:intro_sasrec_perf_seqlen_dist}. The transformers in both models are trained with the length being $100$. The observations are twofold. Firstly, most of the sequences are rather short. Nearly $75\%$ sequences consist of less than $7$ items. Moreover, the performance of very short sequences (\textit{e.g.}, length $= 3$) is much poorer than that of long sequences (\textit{e.g.}, length $\geq 20$). This discrepancy implies the necessity of informative contexts for encoding sequences~\cite{xia2020cg,xia2020composed}, which are limited in short sequences. Therefore, it motivates us to devise \textit{lengthening augmentation for short sequences}.

% There is one most recent work~\cite{xie2020contrastive}, discussing three techniques to augment the training sequences, which are `crop', `mask', and `reorder' operations upon sequences. However, these operations are not applicable to short sequences, because `crop' and `mask' lead to shorter sequences while `reorder' cannot provide extra context information. During the inference period, if we reorder short sequences, their item transition relations are vulnerable to be broken since they contain few items. It is required to augment those short sequences, while still preserving the original sequential correlations.

The challenges for this augmentation are threefold. Firstly, since the target in sequential recommendation is to predict the next item, it is required to maintain the original sequential correlations after augmenting those short sequences. Additionally, though some recent works leverage item attributes~\cite{zhou2020s3,yao2020self} as additional contexts, similar items might not reflect complex items transition 
correlations. Meanwhile, the fact that sequences are the only available data for most SRs further hinders the augmentation for short sequences.

To this end, we propose to augment those short sequences with pseudo-prior items. Intuitively, we append a fabricated sub-sequence of items at the beginning of short sequences, which provide additional contexts~\cite{zhang2020seal,jiao2020sub}. To generate these pseudo-prior items, we pre-train a transformer from a reverse (i.e., right-to-left) direction of the original sequence to predict the prior item. As such, it can preserve the sequential correlations among these pseudo-prior items and the original items in sequences.  Given the augmented sequences, we fine-tune the transformer from the original direction (i.e., left-to-right) to predict the next item in a sequence. We name this framework as \textbf{A}ugmenting \textbf{S}equential \textbf{Re}commendation with \textbf{P}seudo-prior \text{i}tems~(\modelname). The contributions are as follows:
\begin{itemize}
    \item To the best of our knowledge, we are the first work that investigates the possibility of improving the performance of SR by augmenting short sequences with pseudo-prior items.
    \item We design a novel framework \modelname that pre-trains over reverse-direction sequences and fine-tunes on augmented sequences.
    \item The \modelname significantly outperforms existing transformer-based SR models (\textit{e.g.,} 18.1\% for Recall@5). Detailed analyses verify the effectiveness of augmenting short sequences. 
\end{itemize}

\section{Preliminary}
\subsection{Problem Definition}
In SR problem, we denote the user set and item set as $\mathcal{U}$ and $\mathcal{V}$, respectively, whose user and item element are denoted as $u$ and $v$, respectively. Each user is associated with a sequence of items $\mathcal{S}^{u}=[v_{1}^{u}, v_{2}^{u}, \ldots, v_{T_u}^{u}]$, 
% \left|S^{u}\right|}^{u}
which are in chronological order of the interaction time with the user $u$. $T_u=|\mathcal{S}^{u}|$, denoting the total number of items in the sequence of user $u$. The SR problem is commonly evaluated as the next-item prediction, which is formulated as:
\begin{equation}
p\left( v_{T_u+1}^{(u)}=v \left|  \mathcal{S}^{u} \right.\right),
\end{equation}
which is interpreted as calculating the probability of all candidate items, given the training sequence $\mathcal{S}^{u}$.

\subsection{Embedding}
We maintain an embedding table $\mathbf{E}\in \mathbb{R}^{d\times |\mathcal{V}|}$ for all the items, whose elements $\mathbf{e}_i \in \mathbb{R}^{d}$ denotes the embedding for item $v_i$. Besides, we should train embeddings for the position indices in a sequence. Since a transformer is trained with a fixed length sequence, we hold a position embedding table $\mathbf{P}\in \mathbb{R}^{d\times n}$, where $n$ is the maximum length. Its element $\mathbf{p}_i$ denotes the position embedding for $i$-th position in a sequence. For a sequence $\mathcal{S}$, we truncate it to be the last $n$ items if it is longer than $n$. And, we employ zero-padding~\cite{kang2018self} if it is shorter than $n$. Therefore, given a sequence $\mathcal{S}={[v_1,v_2,\dots,v_n]}$, the input embedding is 
\begin{equation}\label{eq:sequence_embed}
    \mathbf{E}_{\mathcal{S}} = [\mathbf{e}_1+\mathbf{p}_1, \mathbf{e}_2+\mathbf{p}_2, \dots, \mathbf{e}_n+\mathbf{p}_n].
\end{equation}

\subsection{Transformer for SR}
We illustrate the basic structure of a transformer~\cite{vaswani2017attention} as in the Figure~\ref{fig:model}(a). It can encode a sequence as in Eq.~(\ref{eq:sequence_embed}) to an output embedding. To be more specific, a transformer consists of two components, \textit{multi-head attention} and \textit{feed-forward network}.
The multi-head attention adopts scaled dot-product attention at each head to learn the importance of items in sequences. The multi-head attention on the input embeddings is:
\begin{align}
     \mathbf{H} &= \text{concat}\{\text{head}_1,\text{head}_2,\dots,\text{head}_h\}\mathbf{W}^{O}, \\
    \text{head}_i &= \text{Attention}(\mathbf{W}_i^{Q}\mathbf{E}_{\mathcal{S}},\mathbf{W}_i^{K}\mathbf{E}_{\mathcal{S}},\mathbf{W}_i^{V}\mathbf{E}_{\mathcal{S}}),
\end{align}
where the \textit{Attention} is the scaled dot-product attention~\cite{vaswani2017attention,kang2018self,sun2019bert4rec}, and its input are query, key and value which are linearly mapped from $\mathbf{E}_{\mathcal{S}}$ with $\mathbf{W}_i^{Q}$, $\mathbf{W}_i^{K}$ and $\mathbf{W}_i^{V}$, respectively. Then, we use a non-linear Feed-Forward Network~(FFN) to position-wisely map $\mathbf{H}_i$ to the sequence embedding at position $i$.
The FFN consists of two linear transformation layers with a ReLU activation function in between~\cite{vaswani2017attention,kang2018self}. 
% It maps $\mathbf{H}$ as:
% \begin{equation}
%     \text{FFN}(\mathbf{H}_i) = \mathbf{W}_{2}\operatorname{ReLU}\left(\mathbf{W}_{1}\mathbf{H}_{i} +\mathbf{b}_{1}\right) +\mathbf{b}_{2}.
% \end{equation}
% where $\mathbf{H}_i$ is the position-wise element of $\mathbf{H}$, which is used to predict the item at position $i$.
A transformer layer is denoted as \transformer. 
% Stacking $L$ \transformer layers models complex item transition correlations. Moreover, residual connections, layer normalization, and drop-out are employed between layers. 
More details about the transformer for SR can be found in~\cite{kang2018self,sun2019bert4rec,xie2020contrastive}.

\section{Proposed Model}
\begin{figure*}
    \centering
    \includegraphics[width=0.8\linewidth]{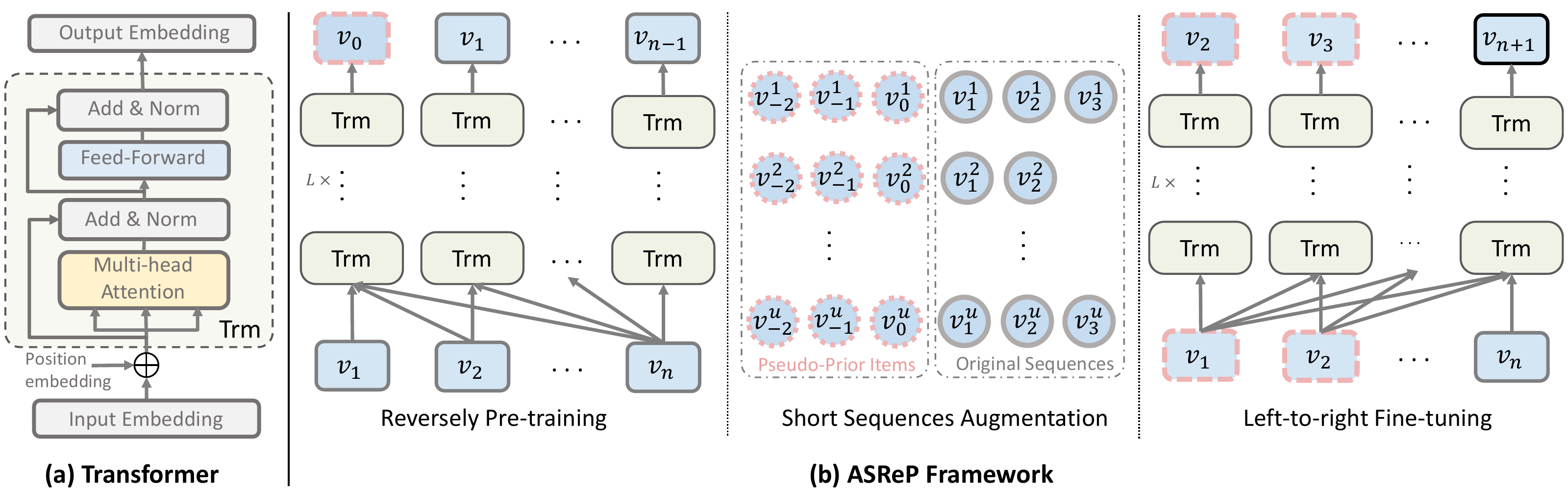}
    \caption{Transformer and \modelname framework. \modelname consists of three components: Firstly, \textit{reversely pre-training} aims to predict prior items (pink dash blocks) of sequences. Secondly, \textit{short sequences augmentation} is to generate $k$ pseudo-prior
    \label{fig:model} items for short sequences ($M=3$), where $k=3$ in this figure. Finally, \textit{left-to-right fine-tuning} is to train the model for next-item ($v_{n+1}$) prediction with augmented sequences. }
\end{figure*}

In this section, we present the framework of \modelname, which consists of \textit{reversely pre-training}, \textit{short sequence augmentation}, and \textit{left-to-right fine-tuning}.  Since this paper's main purpose is to verify the efficacy of augmenting short sequences, we directly adopt the transformer backbone in SASRec~\cite{kang2018self} as the sequence encoder. The framework is illustrated in Figure~\ref{fig:model}(b).

\subsection{Reversely Pre-training}
To generate pseudo-prior items for short sequences, we train the transformer from the right-to-left reverse direction of sequences, which is presented in the left-hand side of Figure~\ref{fig:model}(b). As such, the transformer is able to predict the prior item of a sequence $\mathcal{S}={[v_1,v_2,\dots,v_n]}$ as:
\begin{equation}
    p(v_{0}=v \left| \mathcal{S} \right.),
\end{equation}
where $v_0 \in \mathcal{V}$ denotes the prior item of the sequence. In the implementation, we mask the left prior item and enforce the transformer to predict it, which is exactly the reverse direction of the training in SASRec. Note that, though the model is trained in a reverse direction, the self-attention module can also reveal the item correlations, which is why the previous work~\cite{sun2019bert4rec} endows a sequence encoder with a bi-directional transformer. 

\subsection{Short Sequences Augmentation}
Next, we use the reversely pre-trained transformer to recursively generate pseudo-prior items. It involves two hyper-parameters:
\begin{itemize}
    \item $k$, denoting the number of pseudo-prior items ahead the sequence.
    \item $M$, augmenting a sequence if its length $\leq M$.
\end{itemize}
We denote these pseudo-prior items as $[v_{-k+1}, \dots, v_{-1}, v_{0}]$ and put them in the beginning of the original short sequence $\mathcal{S}$. We illustrate the augmentation for all short sequences as in the middle of Figure~\ref{fig:model}(b). Note that these pseudo-prior items are generated recursively. For example, we generate $v_0$ and append it ahead of $\mathcal{S}$, which constitutes a new sequence for the inference of $v_{-1}$.

\subsection{Left-to-Right Fine-tuning}
Finally, we fine-tune the transformer using the augmented sequences from the left-to-right direction to predict the next item. We illustrate this on the right-hand side of Figure~\ref{fig:model}(b). We use the augmented sequence to predict the next item $v_{n+1}$.  The first $k$ items for short sequences are all pseudo-prior items, which are those pink dash boxes. Long sequences without augmentation have no pseudo-prior items. 
% The given short sequence is augmented with pseudo-prior items, represented as pink dash blocks.
Note that we mask the right next item to maintain the causality and sequential transition correlations for the attention module in \transformer layers. Again, since the transformer requires a fixed length of input, we truncate it to the last item if the sequence (both augmented and without augmentation) is longer than $n$,  and zero-pad it if it is still shorter than $n$.

\section{Experiments}
In this section, we present experimental settings and results. We will answer the following Research Questions~(\textbf{RQs}):
\begin{itemize}
    \item \textbf{RQ1}: Is \modelname effective in improving the performance of sequential recommendation?
    \item \textbf{RQ2}: What is the best choice of the number of pseudo-prior items ($k$) and the length of short sequences ($M$) of \modelname?
    % \item \textbf{RQ3}: Can those generated pseudo-prior items provide extra context information?
    \item \textbf{RQ3}: How does the \modelname perform with respect to the length of original sequences? 
    
\end{itemize}
\subsection{Datasets}
In our experiments, we use two publicly available datasets~\cite{mcauley2015image}: (1) Beauty: Amazon Beauty 5-core includes 22,363 users~(sequences), 12,101 items, and 198,502 ratings with density as 0.07\%. The shortest sequence length is 5 while more than 75\% of sequences are less than 9, and only 1,019 sequences are longer than 20. (2) Phones: Amazon Cell Phones and Accessories
5-core includes 27,879 users~(sequences), 10,429 items, and 194,439 ratings with density as 0.06\%. The shortest sequence length is 5 while more than 75\% of sequences are less than 7, and only 284 sequences are longer than 20.

Following~\cite{kang2018self, sun2019bert4rec}, we transform datasets with explicit ratings into implicit feedbacks by treating the presence of a review as positive feedback, and use timestamps to determine the order of items within each sequence (per user). We use the most recent item of each user for testing and the second most recent item for validation.

\begin{table}[]
\caption{Performance Comparison in Recall@5, NDCG@5, and MRR. The best and second-best results are boldfaced and underlined, respectively.}
\label{tab:overall_perf}
\resizebox{0.47\textwidth}{!}{%
\begin{tabular}{l|ccc|ccc}
\toprule
                    & \multicolumn{3}{c|}{Beauty} & \multicolumn{3}{c}{Phones}       \\
                    & Recall@5  & NDCG@5 & MRR    & Recall@5 & NDCG@5 & MRR          \\ \hline
BPRMF               & 0.3737    & 0.2712 & 0.2682 & 0.3862   & 0.2849 & 0.2831       \\
LightGCN            &   0.3852        &  0.2927     & 0.2906      &   0.4498       &   0.3394     &   0.3218         \\ \hline
SASRec              & 0.3963    & 0.2949 & 0.2907 & 0.4646   & 0.3379 & 0.3314       \\
BERT4Rec            & 0.4143    & 0.3128 & 0.3098 & 0.5077   & 0.3812 & {\ul 0.3720} \\ \hline
ItemCor             & 0.4053    & 0.3067 & 0.3039 &   0.4755       &   0.3526     &  0.3428            \\
re-train                & {\ul 0.4311}    & {\ul 0.3257}    & {\ul 0.3209}    & {\ul 0.5213}    & {\ul 0.3833}    & 0.3667          \\
\textbf{\modelname} & \textbf{0.4684} & \textbf{0.3547} & \textbf{0.3458} & \textbf{0.5573} & \textbf{0.4193} & \textbf{0.4026} \\ \midrule\midrule
\text{\small v.s. SASRec} &   +18.1\%        &  +20.2\%      &  +18.9\%      &   +19.9\%       &  +24.0\%      & +21.4\%\\
\text{\small Improve.} &    +8.6\%       &  +8.9\%      & +7.7\%      &   +6.9\%      &   +9.3\%     &     +8.2\%         \\ \bottomrule
\end{tabular}%
}
\end{table}

\subsection{Experimental Settings}
\textbf{Evaluation Protocols.} We evaluate all models in three metrics: Recall@5, NDCG@5, and Mean Reciprocal Rank~(MRR). Recall@5 measures the fraction of relevant items being retrieved at top-5 recommendations out of all relevant items, NDCG@5 evaluates their top-5 ranking performance, while MRR measures the ranking performance of the entire ranking list. For each user, we randomly sample 100 negative items for ranking
with the ground-truth item. \hfill \break
% To correct the sampling bias for evaluation~\cite{krichene2020sampled}, we apply the unbiased estimator in~\cite{krichene2020sampled} to correct the sampled ranks.
\textbf{Baselines.} We compare \modelname with two transformer-based SR models SASRec~\cite{kang2018self} and BERT4Rec~\cite{sun2019bert4rec}. Also,  we compare with two static models, the BPR-MF~\cite{10.5555/1795114.1795167} and  LightGCN~\cite{10.1145/3397271.3401063}. Additionally, we create two variants of the \modelname. One is \textit{ItemCor}, whose pseudo-prior items are items with similar embeddings from LightGCN. Hence, it augments short sequences without reversely training but only item correlations. The other one is \textit{re-train}, which re-trains a new transformer over the augmented sequences rather than fine-tuning the reversely pre-trained transformer. \\
% \break
\textbf{Parameter Settings.} For all methods, we search the embedding size $d$ from $\{32, 64, 128\}$. The $L2$ regularization term is selected from $\{0.0, 0.0001, 0.001, 0.01, 0.1\}$. For max input sequence length $n$, we search from $\{50, 100\}$ for all SR methods. We search the number of layers $L$ from $\{1,2,3\}$. For BERT4Rec, we also search the masked probability from $\{0.2, 0.3, 0.5, 0.7\}$. We grid search all possible combinations of hyper-parameters on the validation set. The best hyper-parameters set for Beauty dataset is max length $n=100$, $d=128$, $L=2$, dropout rate is $0.7$, $L2=0.0$, short sequence length threshold $M=18$, and the number of augmenting items $k=15$. For Phones dataset, it is max length $n=100$, $d=32$, $L=2$, dropout rate is $0.5$, $L2=0.0$, $M=18$, and $k=17$.

\subsection{Overall Comparison~(RQ1)}
We report the overall performance comparison of \modelname and other baselines in Table~\ref{tab:overall_perf}. The observations are as follows:
\begin{itemize}
    \item \modelname outperforms other baselines on all metrics. Compared with SASRec (v.s. SASRec), \modelname has significant relative improvements on all three metrics, in average  $19.1\%$ and $21.8\%$ on Beauty and Phones, respectively. Compared with the second-best one (improvements are in the last row), it also has in average $8.4\%$ and $8.1\%$ improvements on Beauty and Phones dataset, respectively. These results prove that our framework is very effective in modeling item sequential correlations. And, the augmentation for short sequences is rather crucial.
    \item All sequential models are better than static methods, \textit{i.e.,} BPR-MF and LightGCN, which verifies the efficacy of modeling item sequential correlations with transformer. However, they are also worse than \modelname as they suffer from the cold-start issue when predicting the next item for short sequences. 
    \item The two variants of \modelname perform worse than \modelname. Compared with \modelname, \textit{ItemCor} directly append similar items before short sequences as pseudo-prior items. The worse performance of it shows the necessity of reversely pre-training a transformer for generating pseudo-prior items. But it is still better than SASRec, which proves the benefits of augmenting short sequences contexts. Compared with \modelname, \textit{re-train} re-trains a new transformer. It ignores the reversely pre-trained transformer and trains a new transformer for recommendation. Its worse performance compared with \modelname verifies that pre-training helps capture important sequential item correlations, which is also why BERT4Rec performs better than SASRec. 
\end{itemize}

\subsection{Parameter Sensitivity~(RQ2)}
\begin{figure} %[ht]
     \centering
     \begin{subfigure}[b]{0.22\textwidth}
         \centering
        \includegraphics[width=1\textwidth]{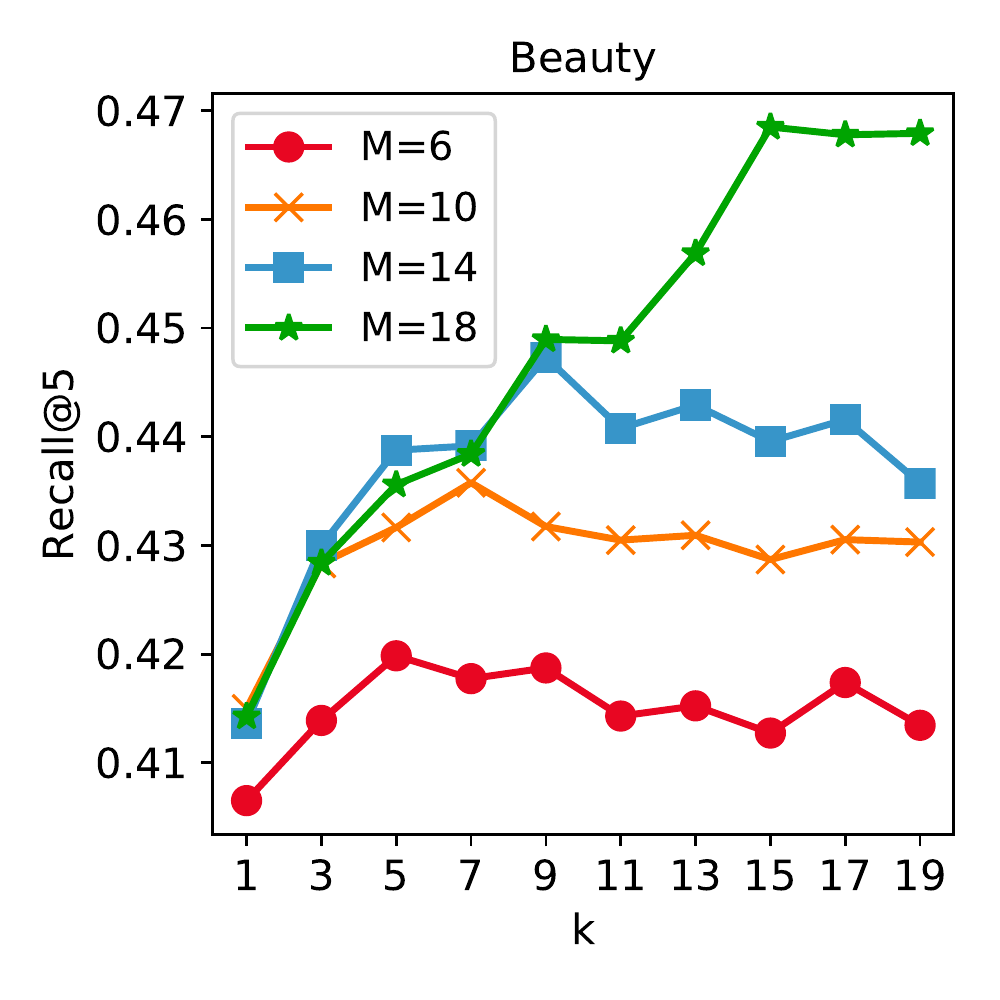}
        \label{fig:beauty_all_recall_params}
     \end{subfigure}
     \begin{subfigure}[b]{0.22\textwidth}
         \centering
        \includegraphics[width=1\textwidth]{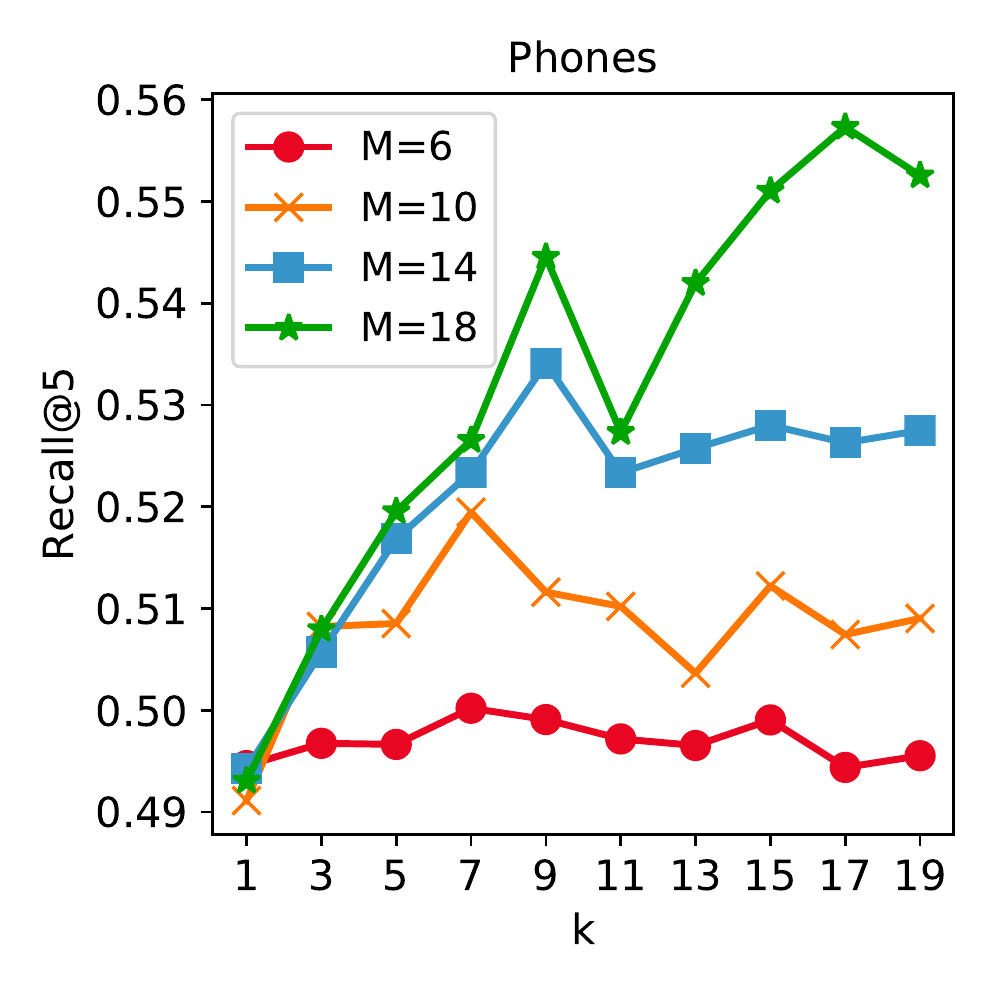}
        \label{fig:phones_all_recall_params}
     \end{subfigure}
    \vspace{-5mm}
     \caption{The Recall@5 performance on different values of $M$ and $k$ over two datasets.}
     \label{fig:perf_wrt_params}
\end{figure}
We analyze the performance of \modelname w.r.t. its two hyper-parameters $k$ and $M$, which denote the number of pseudo-prior items and the threshold for short sequences, respectively. We select $k \in \{1,3,\dots,19\}$ and $M \in \{6,10,14,18\}$. The results are illustrated in Figure~\ref{fig:perf_wrt_params}. On both datasets, the performance improves when $M$ increases, which verifies the importance of augmenting short sequences. Additionally, when $M$ is small, \textit{e.g.} $6$, increasing $k$ has little improvements. This is because those generated pseudo-prior items are not informative when we only consider very short sequences. Though performance improves when we increase both $k$ and $M$, the time cost also increases. Therefore, we should find a trade-off between those hyper-parameters and efficiency. 

\subsection{Performance w.r.t. Sequence Length~(RQ3)}

\begin{figure}[ht]
     \centering
     \begin{subfigure}[b]{0.22\textwidth}
         \centering
        \includegraphics[width=1\textwidth]{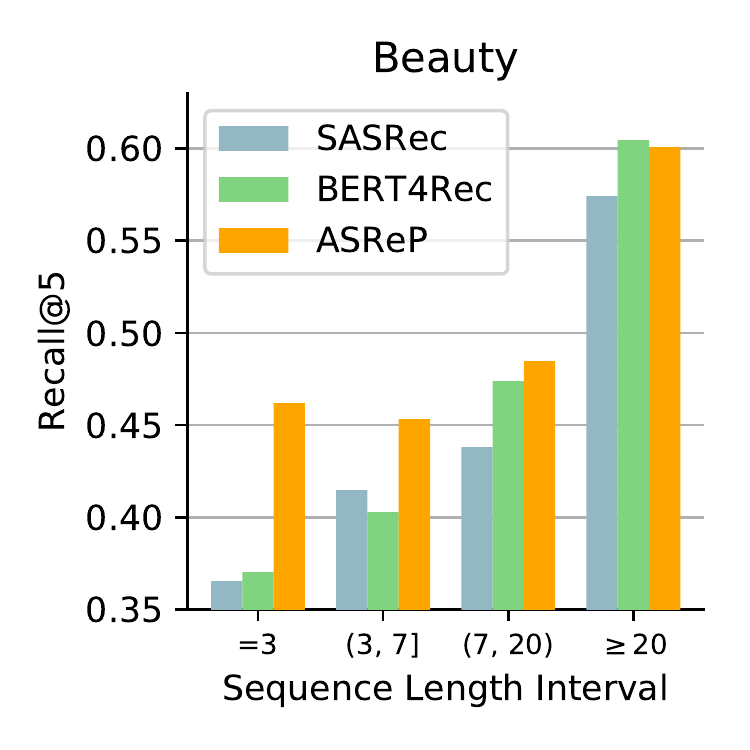}
        \label{fig:beauty_all_recall_seqlength}
     \end{subfigure}
     \begin{subfigure}[b]{0.22\textwidth}
         \centering
        \includegraphics[width=1\textwidth]{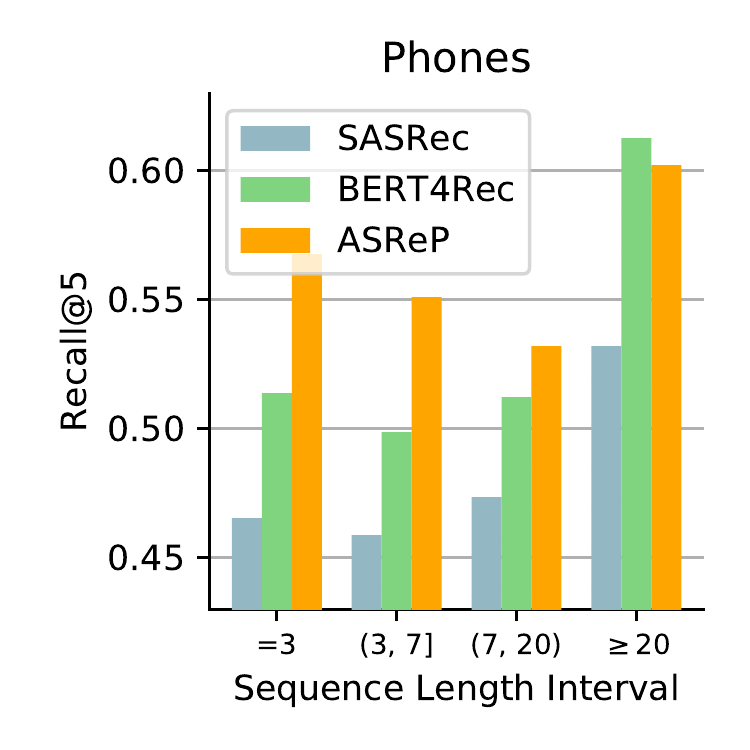}
        \label{fig:phones_all_recall_seqlength}
     \end{subfigure}
     \vspace{-5mm}
     \caption{The Recall@5 performance on different sequence lengths over two datasets.}
     \label{fig:perf_wrt_length}
\end{figure}

We illustrate the performance of SASRec, BERT4Rec and \modelname w.r.t. the sequence length in Figure~\ref{fig:perf_wrt_length}. Firstly, we observe that \modelname significantly outperforms both SASRec and BERT4Rec when sequence length $< 20$, i.e., short sequences. This proves that \modelname is effective in augmenting  short sequences and provides additional contexts to encode short sequences. Secondly, since we directly adopt the same fine-tuning strategy as SASRec, the better performance of \modelname compared with SASRec on all sequence lengths demonstrates the necessity of augmenting short sequences. Additionally, we observe that BERT4Rec performs the best when sequence length $\geq 20$, which shows the benefits of a bi-directional transformer. However, since we focus on augmenting short sequences and most sequences are rather short, \modelname can thus significantly improve the overall performance. It is also worth noting that we could substitute the left-to-right fine-tuning of \modelname to bi-directional fine-tuning, which may also outperform BERT4Rec on long sequences. We leave it for future study. 

\section{Conclusions}
In this paper, we study improving sequential recommendation via augmenting short sequences. To complete this task,  we propose a new framework, \modelname, which employs a reversely pre-trained transformer to generate pseudo-prior items for short sequences. We fine-tune the pre-trained transformer from the left-to-right direction to predict the next item in a sequence. Moreover, we conduct overall comparisons of \modelname with other baselines, verifying the effectiveness of \modelname. Detailed analyses demonstrate that \modelname significantly improves the performance regarding short sequences. 

\section{Acknowledgements}
This work is supported in part by NSF under grants III-1763325, III-1909323, and SaTC-1930941. 

% \usepackage{graphicx}
% \usepackage[normalem]{ulem}
\newpage
\bibliographystyle{ACM-Reference-Format}
\bibliography{sample-base}


\end{document}